\documentclass[prb,preprint]{revtex4-1} 
\usepackage[utf8]{inputenc}
\usepackage{amsmath, natbib, graphicx, enumerate, tikz, float, amsthm, verbatim}
\usepackage[polish, english]{babel}
\usetikzlibrary{patterns}
\usetikzlibrary{arrows.meta}

\newcommand{\dbar}{d\hspace*{-0.08em}\bar{}\hspace*{0.1em}}

\begin{document}

\title{Using the Carnot cycle to determine changes of the phase transition temperature}

\author{Oskar Grocholski}
\email{o.grocholski@student.uw.edu.pl}
\author{Kornel Howil}
\email{k.howil@student.uw.edu.pl}
\author{Stanisław Rakowski}
\email{sl.rakowski@student.uw.edu.pl}
\affiliation{University of Warsaw, Faculty of Physics, Pasteura 5, 02-093 Warsaw, Poland}

\author{Piotr Maksymiuk}
\email{2003piotr@gmail.com}

\affiliation{Polish Children's Fund, Pasteura 7, 02-093 Warsaw, Poland}
 
\date{\today}

\begin{abstract}
The Clausius-Clapeyron relation and its analogs in other first-order phase transitions, such as type-I
superconductors, are derived using very elementary methods, without appealing to the more advanced
concepts of entropy or Gibbs free energy. The reasoning is based on Kelvin's formulation of the second
law of thermodynamics, and should be accessible to high school students. After recalling some basic facts about the Carnot cycle, we present two very different
systems that undergo discontinuous phase transitions (ice/water and normal/superconductor), and
construct engines that exploit the properties of these systems to produce work. In each case, we show that
if the transition temperature $T_{\mathrm{tr}}$ were independent of other parameters, such as pressure or magnetic field,
it would be possible to violate Kelvin's principle, i.e., to construct a perpetuum mobile of the second kind.
Since the proposed cyclic processes can be realized reversibly in the limit of infinitesimal changes in
temperature, their efficiencies must be equal to that of an ordinary Carnot cycle. We immediately obtain
an equation of the form $dT/dX=f(T, X)$, which governs how the transition temperature changes
with the parameter $X$.
\end{abstract}

\maketitle

\section{Introduction}
In 1844, Lord Kelvin studied a cyclic process based on the water-ice phase transition, and
concluded that a perpetuum mobile of the second kind would be possible, unless the melting
temperature of ice varied with pressure\cite{James-Thomson}. This variation is described quantitatively by the
Clausius-Clapeyron equation. Usually, it is derived by relying on the continuity of the Gibbs free
energy (or, more precisely, the chemical potential) between the two phases (see Appendix \ref{Appendix-A}).
In this work, we show how to derive the Clausius-Clapeyron equation, and its analogs in other
first order phase transitions, by constructing reversible cycles operating between two very close
temperatures. 

First, let us briefly recall some basic facts about the second law of thermodynamics. In
Kelvin's formulation, it reads, ``It is impossible to devise an engine which, working in a cycle,
shall produce no effect other than the extraction of heat from a reservoir and performance
of an equal amount of mechanical work" \cite{Pippard}. In other words, it is impossible to construct a
heat engine that would have 100\% efficiency, defined as the ratio of the work done to the heat
absorbed. Furthermore, this implies that of all possible engines operating between two heat
reservoirs at different fixed temperatures, the ones operating reversibly have the highest
efficiency. This is shown by making the \textit{reduction ad absurdum} argument presented below: if there were a cycle
more efficient than the reversible one, one could construct a cycle violating the second law
of thermodynamics \cite{AJP-reversible}.

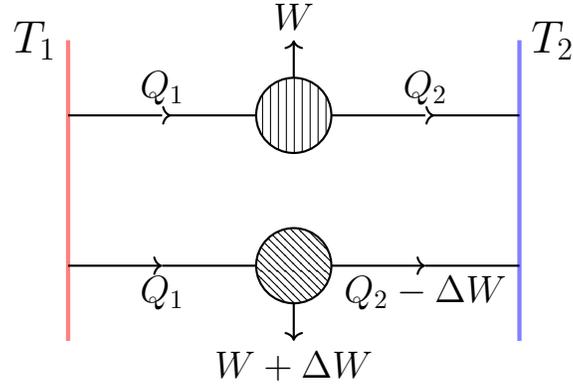
\begin{figure}
        \centering
        \begin{tikzpicture}[scale = 0.5]

\draw[red!50!white, ultra thick] (-6,-4) -- (-6,4) node[black] [anchor=east]{\Large{$T_1$}};
\draw[blue!50!white,ultra thick] (6,-4) -- (6,4) node[black] [anchor=west]{\Large{$T_2$}};
\draw[black, thick, pattern=north west lines](0,-2) circle (1);

\draw[black, thick, pattern=vertical lines](0,2) circle (1);
\draw[black,thick, ->] (-6,-2) -- (-3.5,-2) node[black] [anchor=north]{\large{$Q_1$}};
\draw[black,thick] (-3.5,-2) -- (-1,-2);
\draw[black,thick, ->] (1,-2) -- (3.5,-2)node[black] [anchor=north]{\large{$Q_2-\Delta W$}};
\draw[black,thick] (3.5,-2) -- (6,-2);
\draw[black,thick] (-6,2) -- (-3.5,2);
\draw[black,thick, -<](-1,2) -- (-3.5,2)node[black] [anchor=south]{\large{$Q_1$}};
\draw[black,thick] (1,2) -- (3.5,2);
\draw[black,thick, -<](6,2) -- (3.5,2)node[black] [anchor=south]{\large{$Q_2$}};
\draw[black,thick, ->](0,3) -- (0,4) node[black][anchor=south]{\large{$W$}};
\draw[black,thick, ->](0,-3) -- (0,-4) node[black][anchor=north]{\large{$W+\Delta W$}};
\end{tikzpicture}
\caption{A hypothetical compound cycle consists of two engines. The upper one is reversible while
the lower (hypothetical) one is assumed to be more efficient.}
\label{fig1}
\end{figure}

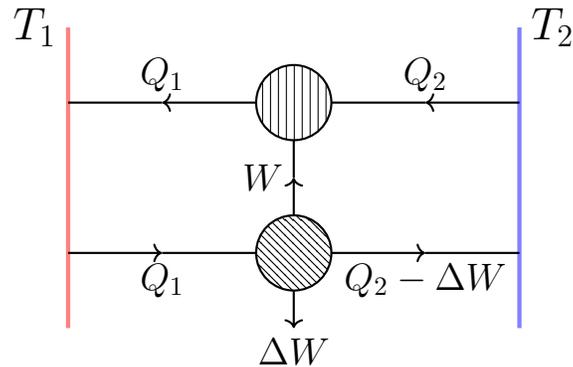
\begin{figure}
        \centering
        \begin{tikzpicture}[scale = 0.5]

\draw[red!50!white, ultra thick] (-6,-4) -- (-6,4) node[black] [anchor=east]{\Large{$T_1$}};
\draw[blue!50!white,ultra thick] (6,-4) -- (6,4) node[black] [anchor=west]{\Large{$T_2$}};
\draw[black, thick, pattern=north west lines](0,-2) circle (1);

\draw[black, thick, pattern=vertical lines](0,2) circle (1);
\draw[black,thick, ->] (-6,-2) -- (-3.5,-2) node[black] [anchor=north]{\large{$Q_1$}};
\draw[black,thick] (-3.5,-2) -- (-1,-2);
\draw[black,thick, ->] (1,-2) -- (3.5,-2)node[black] [anchor=north]{\large{$Q_2-\Delta W$}};
\draw[black,thick] (3.5,-2) -- (6,-2);
\draw[black,thick] (-6,2) -- (-3.5,2);
\draw[black,thick, ->](-1,2) -- (-3.5,2)node[black] [anchor=south]{\large{$Q_1$}};
\draw[black,thick] (1,2) -- (3.5,2);
\draw[black,thick, ->](6,2) -- (3.5,2)node[black] [anchor=south]{\large{$Q_2$}};
\draw[black,thick, ->](0,-3) -- (0,-4) node[black][anchor=north]{\large{$\Delta W$}};
\draw[black,thick, ->](0,-1) -- (0,0) node[black] [anchor=east]{\large{$W$}};
\draw[black,thick](0,0) -- (0,1);
\end{tikzpicture}
\caption{Proof that the reversible cycle has the maximal possible efficiency. Reversing the sense of
operation of the upper (reversible) engine would lead to violation of the second law of thermodynamics.}
\label{fig2}
\end{figure}

Consider the composite cyclic process shown in Fig. \ref{fig1}, which consists of two separate cycles
(represented by hatched circles), which we will also call engines; the upper one is reversible, and
the lower one is a hypothetical cycle that is more efficient than the first. Vertical lines marked $T_1$
and $T_2$ represent two heat reservoirs at the indicated temperatures, and lines with arrows
represent transfers of heat or work in the directions indicated by the arrows. Both of these
engines extract heat $Q_1$ from the left reservoir at temperature $T_1$. Assume that the upper
(reversible) cycle releases heat $Q_2$ to the right reservoir at temperature $T_2$ ($T_2<T_1$) and performs
an amount of work $W=Q_1-Q_2$. Since the lower cycle is more efficient, it releases heat $Q_2 - \Delta W$, and performs work $W+ \Delta W$ ($\Delta W > 0$).
If this were possible, the upper (reversible) engine could be run in the opposite
direction to extract the same heat $Q_2$ from the reservoir at $T_2$, using an amount of work $W$ drawn
from the work $W + \Delta W$ performed by the lower engine (Fig. \ref{fig2}). The overall result would then
be a net amount of heat $\Delta W$ taken from the reservoir at $T_2$ converted entirely into positive work
$\Delta W$, in contradiction to Kelvin's assertion.

From the reasoning summarized in Figs \ref{fig1} and \ref{fig2},
it also follows that \textit{all} reversible cycles working between the same two reservoirs must have the same efficiency, and that their efficiency depends only on these two temperatures and nothing else. This efficiency is shown in any textbook\cite{Pippard} to be given by the Carnot formula:
\begin{equation}
    \eta = 1 - \frac{Q_{out}}{Q_{in}} = 1 - \frac{T_2}{T_1} = \frac{T_1 - T_2}{T_1}~\!.
    \label{eq-Carnot}
\end{equation}
Note that the efficiency of a reversible cycle decreases to zero (no work can be
obtained) in the limit $T_2\rightarrow T_1$.
This result will be the main tool in the analyses
presented in the remaining part of this paper.

\section{A simple derivation of the Clausius-Clapeyron relation}\label{sec-II}
In 1844, while analyzing a cycle in which freezing water performed mechanical work, Lord
Kelvin was led to a paradox: it appeared that by exploiting the ice-water phase transition, it was
possible to construct a cycle violating the second law of thermodynamics \cite{Kelvin-paradox}. To avoid this,
Kelvin's brother James concluded that the ice melting temperature must decrease with increasing
pressure \cite{James-Thomson}. Here we show how, with the help of a suitable infinitesimal cycle, this qualitative
statement can be converted into a precise mathematical formula (the Clausius-Clapeyron
equation).

In the water-ice engine considered by Kelvin (illustrated in Fig. \ref{cycle1}), initially a movable deck forms the upper wall of a container filled with water at a temperature $T_1$ that is greater than its freezing point $T_0$ (step 1).  In step 2, a mass $M$ is placed on the deck.  Owing to the very small compressibility of water, the change in water volume from step 1 to step 2 will be ignored in this analysis; in the limit of an infinitesimal cycle this can be justified.  In step 3, the system is brought into contact with a cold reservoir at a temperature $T_2 < T_0$ and the water freezes. Because the density of ice is less than that of water, the body is lifted up by $ \Delta H =  (V_{\rm ice} - V_{\rm water}) / A $, where $V_{\rm water/ice}$ is the volume of the water/ice and $A$ is the cross-sectional area of the container. In step 4, the mass is removed from the deck and the gain in its potential energy $ Mg\Delta H $ is transformed into a mechanical work by lowering $M$ to its initial height. Finally, the engine
absorbs heat from a warmer reservoir $T_1 > T_0$ and the ice melts. The heat received by the engine
from the hotter reservoir is equal to the sum of the latent heat of melting of the ice plus the heat
needed, first to bring the ice to the melting temperature, and then to warm the resulting water to
the temperature of the warmer reservoir. The two last heats can be neglected if the cycle is
infinitesimal. After this sequence of operations,  the engine returns to its initial state, which was before $M$ was placed on it. Figure \ref{phase-diagram1} shows this cycle on the water-ice phase diagram using T-p variables.
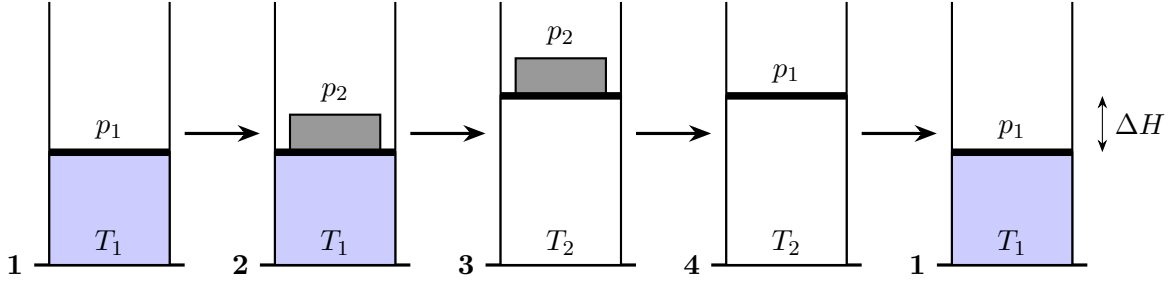
\begin{figure}
\centering
\begin{tikzpicture}
    \draw[black, very thick] (0,0)node[left]{\textbf{1}} -- (2,0);
    \draw[black, thick] (0.2, 0) -- (0.2, 3.5);
    \draw[black, thick] (1.8, 0) -- (1.8, 3.5);
    \filldraw[fill=blue!20!white, thick] (0.2, 0) rectangle (1.8, 1.5);
    \draw[black, line width=1mm] (0.2, 1.5) -- (1.8, 1.5);
    \node at (1, 0.3) {$T_1$};
    \node at (1, 1.8) {$p_1$};
    
    \draw[black, very thick, -{Stealth}] (2, 1.75) -- (3, 1.75);

    \draw[black, very thick] (3,0)node[left]{\textbf{2}} -- (5,0);
    \draw[black, thick] (3.2, 0) -- (3.2, 3.5);
    \draw[black, thick] (4.8, 0) -- (4.8, 3.5);
    \filldraw[fill=blue!20!white, thick] (3.2, 0) rectangle (4.8, 1.5);
    \filldraw[fill=black!40, thick] (3.4, 1.5) rectangle (4.6, 2);
    \draw[black, line width=1mm] (3.2, 1.5) -- (4.8, 1.5);
    \node at (4, 2.3) {$p_2$};
    \node at (4, 0.3) {$T_1$};

    \draw[black, very thick, -{Stealth}] (5, 1.75) -- (6, 1.75);
    
    \draw[black, very thick] (6,0)node[left]{\textbf{3}} -- (8,0);
    \draw[black, thick] (6.2, 0) -- (6.2, 3.5);
    \draw[black, thick] (7.8, 0) -- (7.8, 3.5);
    \filldraw[fill=white, thick] (6.2, 0) rectangle (7.8, 2.25);
    \filldraw[fill=black!40, thick] (6.4, 2.25) rectangle (7.6, 2.75);
    \draw[black, line width=1mm] (6.2, 2.25) -- (7.8, 2.25);
    \node at (7, 3.05) {$p_2$};
    \node at (7, 0.3) {$T_2$};
    
    \draw[black, very thick, -{Stealth}] (8, 1.75) -- (9, 1.75);
    
    \draw[black, very thick] (9,0)node[left]{\textbf{4}} -- (11,0);
    \draw[black, thick] (9.2, 0) -- (9.2, 3.5);
    \draw[black, thick] (10.8, 0) -- (10.8, 3.5);
    \filldraw[fill=white, thick] (9.2, 0) rectangle (10.8, 2.25);
    \draw[black, line width=1mm] (9.2, 2.25) -- (10.8, 2.25);
    \node at (10, 2.55) {$p_1$};
    \node at (10, 0.3) {$T_2$};
    
    \draw[black, very thick, -{Stealth}] (11, 1.75) -- (12, 1.75);
    
    \draw[black, very thick] (12,0)node[left]{\textbf{1}} -- (14,0);
    \draw[black, thick] (12.2, 0) -- (12.2, 3.5);
    \draw[black, thick] (13.8, 0) -- (13.8, 3.5);
    \filldraw[fill=blue!20!white, thick] (12.2, 0) rectangle (13.8, 1.5);
    \draw[black, line width=1mm] (12.2, 1.5) -- (13.8, 1.5);
    \node at (13, 1.8) {$p_1$};
    \node at (13, 0.3) {$T_1$};
    \draw [{Stealth}-{Stealth}] (14.2, 1.5) -- (14.2, 2.25);
    \node at (14.7, 1.875) {$\Delta H$};
\end{tikzpicture}
    \caption{Kelvin's water-ice engine. Blue color (online) indicates that the water is in the liquid state,
and white color means that it is frozen. $p_1$ is the external (atmospheric) pressure. $p_2$ is the total
pressure exerted on the water/ice after the weight has been placed on the deck.}
    \label{cycle1}
\end{figure}
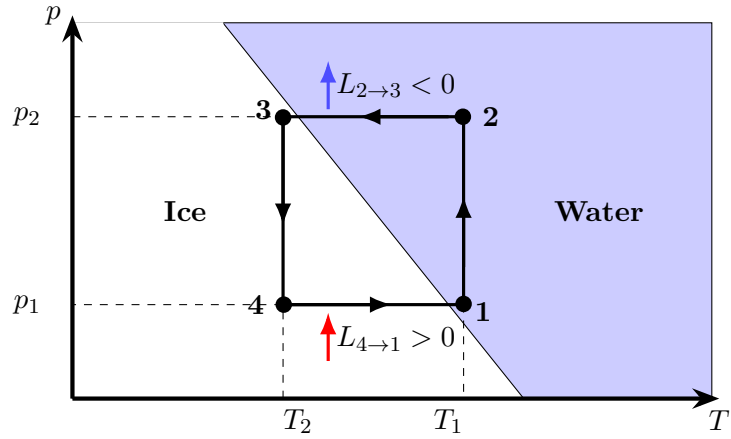
\begin{figure}
     \centering
    \begin{tikzpicture}
    \filldraw[fill=blue!20!white, draw=black] (0,0) rectangle (8.5,5);
    \draw[fill=white] plot[smooth,samples=100,domain=2:6, variable=\x] (\x,{15/2-5*\x/4}) -- plot[smooth,samples=100,domain=8.5:2, variable=\x] (\x,{0});
    \filldraw[fill=white, draw=white] (0,0) rectangle (2,5);
    \draw (0,0) -- (0,4);
    \draw[-{Stealth},ultra thick] (0,0)--(0,5.1) node[pos =1, left] {$p$};
    \draw[-{Stealth},ultra thick] (0,0)--(8.6,0) node[pos =1, below] {$T$};
    \node at (1.5, 2.5) {\textbf{Ice}};
    \node at (7, 2.5) {\textbf{Water}};
    \node at (-0.6, 3.75) {$p_2$};
    \node at (-0.6, 1.25) {$p_1$};
    \node at (3, -0.3) {$T_2$};
    \node at (5, -0.3) {$T_1$};

    \draw[-{Circle}, very thick] (2.8,3.75) -- (5.3, 3.75) node [above, right]{\textbf{2}};
    \draw[-{Latex}, very thick] (5,3.75) -- (3.8, 3.75);
    \draw[-{Circle}, very thick] (5.2,3.75) -- (5.2, 1.15) node [above, right]{\textbf{1}};
    \draw[-{Latex}, very thick] (5.2,1.75) -- (5.2, 2.7);
    \draw[-{Circle}, very thick] (5.2,1.25) -- (2.7, 1.25) node [above, left]{\textbf{4}};
    \draw[-{Latex}, very thick] (2.8,1.25) -- (4.25, 1.25);
    \draw[-{Circle}, very thick] (2.8,1.25) -- (2.8, 3.85) node [above, left]{\textbf{3}};
    \draw[-{Latex}, very thick] (2.8,3.75) -- (2.8, 2.3);
    \draw[dashed] (0,3.75) -- (3, 3.75);
    \draw[dashed] (5.2,1.25) -- (5.2, 0);
    \draw[dashed] (3,1.25) -- (0, 1.25);
    \draw[dashed] (2.8,0) -- (2.8, 1.25);
    
    \draw[-{Latex}, very thick, red] (3.4,0.5) -- (3.4, 1.15 );
    \node at (4.3, 0.8) {$L_{4\rightarrow1}>0$};
    
    \draw[-{Latex}, very thick, blue!70!white] (3.4,3.85) -- (3.4, 4.5);
   \node at (4.3, 4.175) {$L_{2\rightarrow3}<0$};
    
\end{tikzpicture}
    \caption{The discussed cycle shown on the phase diagram of water (not to scale). In steps $1\rightarrow2$ and $3\rightarrow4$, a body is placed on or removed from the deck, which results in a change of the pressure. In step $2\rightarrow3$ the water is cooled so that it freezes ($L_{2\rightarrow3}$ is the heat released). In step $4\rightarrow1$ the ice is heated and it melts ($L_{4\rightarrow1}$ is the heat absorbed). If the melting temperature $T_0$ does not depend on the pressure $p$, then the cycle can be executed with reservoir temperatures $T_1$ and $T_2$ arbitrarily close to one another, leading to violation of the second law of thermodynamics.}
    \label{phase-diagram1}
\end{figure}
\noindent The efficiency of the cycle shown in Figs. \ref{cycle1} and \ref{phase-diagram1} is
\begin{equation}
    \eta = \frac{(V_{\rm ice} - V_{\rm water})\frac{Mg}{A}}{L}~\!,
\end{equation}
or
\begin{equation}
    \eta = \frac{Mg}{A} \frac{\Delta v }{l}~\!,
\end{equation}
where $\Delta v$ is the difference between the ice and water molar volumes, and $l$ is the molar
latent heat of the water-ice transition.

As it stands, the efficiency of the constructed engine does not depend on the temperatures of the reservoirs absorbing heat from or supplying heat to the system. In contrast, the efficiency of the Carnot engine approaches zero as $T_1 - T_2 \rightarrow 0$.  This situation violates the second law of thermodynamics, since as the temperatures are brought together, at some point the efficiency of the constructed engine will exceed that of a Carnot engine operating between the same temperatures.

The apparent contradiction is removed if the temperature $T_0$ of the phase transition varies with
the pressure $p$. The change in pressure exerted on the water/ice results from placing
and subsequently removing the weight lifted by the water. This difference in pressure is simply
\begin{equation}
    \Delta p = \pm \frac{Mg}{A}~\!.
    \label{eq-before-discussion}
\end{equation}
We can already conclude (without any quantitative analysis) that the melting temperature must
grow with lowering the pressure. If it fell, then it would be possible to melt ice after removing
the weight by using the same reservoir that previously was used to freeze the water, and no
difference in temperatures would be needed to construct a working cycle. It is a consequence of
the second law of thermodynamics that an increase of melting temperature with decreasing
pressure must accompany the unusual property of water that its volume is larger in the solid
phase.

It is, however, possible to go further and obtain the precise relation between the
changes in the melting temperature and pressure by comparing the efficiency of the
water-ice engine with that of a Carnot cycle. But first it is necessary to ask if the cycle can be
realized (at least in principle) reversibly. If friction is eliminated (a standard assumption in
thermodynamical reasoning), then placing and removing $M$ becomes reversible. On the other
hand, a transfer of heat from a hotter to a colder body is in general an irreversible process.
However, if $M$ is very small, the change in pressure will also be small, so the change in the
freezing point will also be very small (infinitesimal). In this limit, the difference in temperatures
of the two heat reservoirs can be made arbitrarily small, thereby making the requisite heat
transfers reversible, too. Thus, the infinitesimal cycle can be considered reversible and the
comparison of its efficiency with that of the Carnot cycle is justified. It yields
\begin{equation}
    \eta =\frac{Mg}{A} \frac{\Delta v }{l} = -\Delta p \frac{\Delta v }{l} = \frac{\Delta T}{T} ~\!.
\end{equation}
Recasting this result into the equation for $\Delta p/ \Delta T$ we get
\begin{equation}
    \frac{\Delta p}{\Delta T}|_{\substack{\Delta T \rightarrow 0 \\ \Delta p \rightarrow 0}} = -\frac{l}{T\Delta v}~\!,
\end{equation}
which is the well-known Clausius-Clapeyron relation.
A similar derivation can be found on p. 53 of Pippard's textbook \cite{Pippard}.

\section{Engine based on superconductivity}
To show that the approach used above is quite general, we next apply it to the normal
conductor-superconductor phase transition. One characteristic feature of Type-I superconductors
(e.g. metals like tin, indium, aluminum) is the vanishing of their electrical resistance. A second
feature is the expulsion of an applied magnetic field from their interior, called the Meissner
effect. The latter feature will be crucial for the engine that we will discuss, because it is
responsible for a force that pushes a superconducting specimen from a region of stronger
magnetic field in the direction of weaker field. The work $W$ done by this force to slowly
(reversibly) displace a superconducting body of volume $V$ from a region of (locally) uniform
constant field $B_1$ to a region of uniform field $B_2$ equals
\begin{equation}
    W = \frac{1}{2\mu_0} V (B_1^2 - B_2^2)~\!,
    \label{eq-7}
\end{equation}
where $\mu_0$ is the vacuum permeability. Type-I superconductors are characterized by a critical
magnetic field strength $B_c$, above which the superconducting state is abruptly destroyed; i.e., the
material undergoes a discontinuous phase transition. Analogous to the water freezing point, $B_c$
depends on the temperature: the material remains superconducting
only if $B<B_c(T)$.

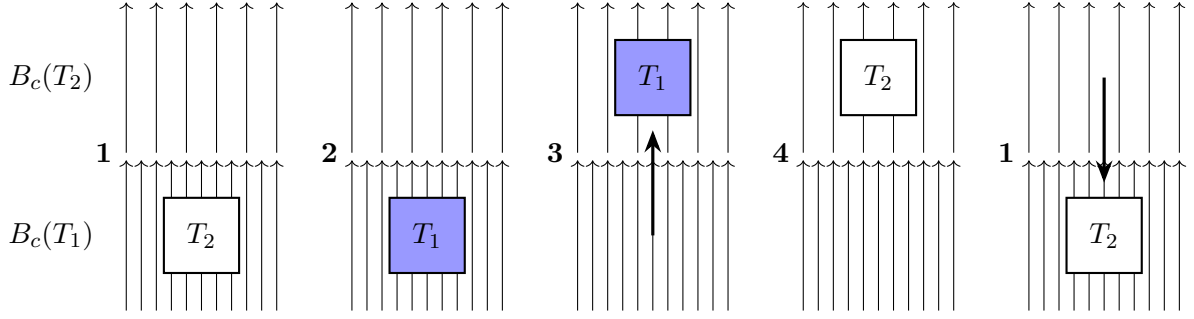
\begin{figure}
\centering
\begin{tikzpicture}

    \foreach \x in {0, 0.4, ..., 2}
    \draw[->,thin] (\x , 0) -- (\x , 2);
    \foreach \x in {3, 3.4, ..., 5}
    \draw[->,thin] (\x , 0) -- (\x , 2);
    \foreach \x in {6, 6.4, ..., 8}
   \draw[->,thin] (\x , 0) -- (\x , 2);
   \foreach \x in {9, 9.4, ..., 11}
   \draw[->,very thin] (\x , 0) -- (\x , 2);
   \foreach \x in {12, 12.4, ..., 14}
   \draw[->,very thin] (\x , 0) -- (\x , 2);
   
   \foreach \x in {0, 0.2, ..., 2}
   \draw[->,thin] (\x , -2.1) -- (\x , -0.1);
   \foreach \x in {3, 3.2, ..., 5}
   \draw[->,thin] (\x , -2.1) -- (\x , -0.1);
   \foreach \x in {6, 6.2, ..., 8}
   \draw[->,thin] (\x , -2.1) -- (\x , -0.1);
    \foreach \x in {9, 9.2, ..., 11}
   \draw[->,thin] (\x , -2.1) -- (\x , -0.1);
   \foreach \x in {12, 12.2, ..., 14}
   \draw[->,thin] (\x , -2.1) -- (\x , -0.1);
   
    \filldraw[fill=white, draw=black, thick] (0.5,-0.6) rectangle (1.5,-1.6);
    \filldraw[fill=blue!40!white, draw=black, thick] (3.5,-0.6) rectangle (4.5,-1.6);
    \filldraw[fill=blue!40!white, draw=black, thick] (6.5,0.5) rectangle (7.5,1.5);
    \filldraw[fill=white, draw=black, thick] (9.5,0.5) rectangle (10.5,1.5);
    \filldraw[fill=white, draw=black, thick] (12.5,-0.6) rectangle (13.5,-1.6);
    
    \draw[-{Stealth}, very thick] (7,-1.1) -- (7,0.3);
    \draw[-{Stealth}, very thick] (13,1) -- (13,-0.4);
    
    \node at (-1, 1) {$B_c(T_2)$};
    \node at (-1, -1.1) {$B_c(T_1)$};
    
    \node at (1, -1.1) {$T_2$};
    \node at (4, -1.1) {$T_1$};
    \node at (7, 1) {$T_1$};
    \node at (10, 1) {$T_2$};
    \node at (13, -1.1) {$T_2$};
    
    \node at (-0.3,0 ) {\textbf{1}};
    \node at (2.7, 0) {\textbf{2}};
    \node at (5.7, 0) {\textbf{3}};
    \node at (8.7, 0) {\textbf{4}};
    \node at (11.7, 0) {\textbf{1}};

\end{tikzpicture}
\caption{
A cyclic process exploiting the phase transition to the superconducting state. Denser lines
indicate the region of stronger magnetic field. The blue (online) color indicates that the body is
in the superconducting state and is not penetrated by the magnetic field. The white color means
that the body is in the normal state and has no magnetic properties.}
\label{superconductor}
\end{figure}

\begin{figure}
    \centering
    \begin{tikzpicture}
    \filldraw[fill=white, draw=black] (0,0) rectangle (8.5,5);
    \draw[fill=blue!40!white] plot[smooth,samples=100,domain=0:8] (\x,{4 - \x * \x / 16}) -- 
    plot[smooth,samples=100,domain=8:0] (\x,{0});
    \draw (0,0) -- (0,4);
    \draw[-{Stealth},ultra thick] (0,0)--(0,5.1) node[pos =1, left] {$B$};
    \draw[-{Stealth},ultra thick] (0,0)--(8.6,0) node[pos =1, below] {$T$};
    \node at (4, 0.5) {\textbf{Superconductor}};
    \node at (7.2, 4) {\textbf{Normal}};
    \node at (-0.6, 3.65) {$B_c(T_1)$};
    \node at (-0.6, 1.75) {$B_c(T_2)$};
    \node at (2, -0.3) {$T_1$};
    \node at (6.2, -0.3) {$T_2$};
    
    \draw[dashed] (0,3.65) -- (2, 3.65);
    \draw[dashed] (6.2,1.75) -- (6.2, 0);
    \draw[dashed] (2,1.75) -- (0, 1.75);
    \draw[dashed] (2,0) -- (2, 1.75);

    \draw[-{Circle}, very thick] (2,3.65)node[left]{\textbf{2}}-- (6.3, 3.65);
    \draw[-{Latex}, very thick] (6.2,3.65) -- (3.8, 3.65);
    \draw[-{Circle}, very thick] (6.2,3.65)node[right]{\textbf{1}} -- (6.2, 1.65);
    \draw[-{Latex}, very thick] (6.2,1.75) -- (6.2, 3);
    \draw[-{Circle}, very thick] (6.2,1.75)node[right]{\textbf{4}} -- (1.9, 1.75);
    \draw[-{Latex}, very thick] (2,1.75) -- (4.2, 1.75);
    \draw[-{Circle}, very thick] (2,1.75)node[left]{\textbf{3}} -- (2, 3.75);
    \draw[-{Latex}, very thick] (2,3.65) -- (2, 2.5);

   % \draw[-{Latex}, very thick, red] (4,1) -- (4, 1.65);
 %   \node at (4.75, 1.35) {$L_{2}>0$};
    
 %   \draw[-{Latex}, very thick, blue!70!white] (4,3.85) -- (4, 4.5);
  %  \node at (4.75, 4.175) {$L_{1}<0$};
\end{tikzpicture}
    \caption{The proposed cycle shown on the phase diagram of the material. Moving the body at temperature $T_2$ to the region of stronger magnetic field is analogous to placing a weight on water in the previous example -- it decreases the temperature of the phase transition.}
    \label{phase-diagram2}
\end{figure}
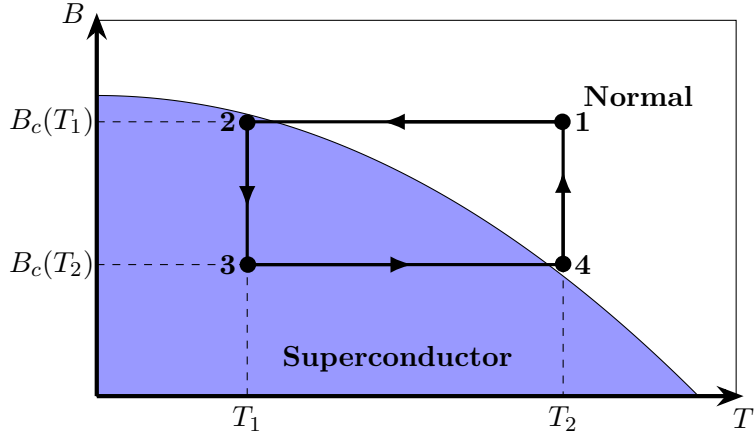

Let us consider the cyclic process presented in Fig. \ref{superconductor}, operating between temperatures $T_1$ and $T_2$, and magnetic fields $B_c(T_1)$ and $B_c(T_2)$, where $T_2 > T_1$. A body, initially in its normal state at
temperature $T_2$ and immersed in field $B_c(T_1)$, is cooled to $T_1$ to enter the superconducting state.
Mechanical work is extracted from the system by allowing the body to be slowly (i.e. reversibly)
pushed into the region of the weaker field $B_c(T_2)$. Next, the body is heated to temperature $T_2$ and
ceases to be superconducting. Finally, the body is returned to the region of the stronger field. No
work is needed to perform this step, because in the normal state, the body does not interact with
the field. This cycle is shown in Fig. \ref{phase-diagram2}, which is a phase diagram of the superconducting
material.
Just as for the water-ice engine, we assume that the cycle is infinitesimal ($T_2 - T_1 = \Delta T \rightarrow 0$), which allows us to neglect the heat received by or extracted from the body other than the latent
heat $L$ related to the phase transition. Therefore, during the cycle the body absorbs a heat $L$ and
performs the work W given by Eqn. \ref{eq-7}, so its efficiency is
\begin{equation}
    \eta=\frac{\frac{1}{2}\frac{V}{\mu_0}\big(B_c^2(T_1)-B_c^2(T_2)\big)}{L}~\!.
    \label{eq-eta-B}
\end{equation}
If the transition temperature were independent of the magnetic field (so that the phase boundary
in Fig. \ref{phase-diagram2} would be a vertical line), one could choose arbitrary values of $B_c(T_{1/2})$, and in the limit $T_1 \rightarrow T_2$ violate the second law of thermodynamics, analogous to the previous example.
Conversely, if the magnetic field at which the transition occurs were independent of temperature,
a working cycle simply could not be constructed, and the comparison with the Carnot cycle
could not be carried out.\\
In the limit of infinitesimal $\Delta B$ and $\Delta T$, Eqn. \eqref{eq-eta-B} can be rewritten as
\begin{equation}
    \eta = -\frac{VB\Delta B}{L\mu_0}~\!,
    \label{eq-eta-BL}
\end{equation}
where $\Delta B = B_c(T_2)-B_c(T_1)<0$.
In the limit $\Delta T, \Delta B \rightarrow 0$ the efficiency given by \eqref{eq-eta-BL} should be equal to the efficiency of the Carnot engine by the same arguments presented in
Section \ref{sec-II}. Hence:
\begin{equation}\label{eq-clap-clau-magnetic}
     \frac{\Delta B}{\Delta T}=-\frac{L\mu_0}{VBT}~\!.
\end{equation}
This the analog of the standard Clausius-Clapeyron relation derived in Section \ref{sec-II}. Eqn. \eqref{eq-clap-clau-magnetic} is usually used to predict the latent heat of the normal conductor-superconductor phase transition: \cite{Saxena}
\begin{equation}
    L=-BT\frac{V}{\mu_0}\frac{\Delta B}{\Delta T}~\!,
\end{equation}
(where $\Delta B/\Delta T < 0$).

\section{Conclusions}
The method presented in this work provides a simple, essentially graphical way of studying
the dependence of the phase transition temperature on external parameters. We have shown
how to apply it using two examples of first-order phase transitions, but there are many other
phase transitions that could be analyzed analogously (e.g. melting, boiling, (re)sublimation, etc.).
In each case, the strategy is the same: identify a discontinuous change of the system that can be
used to perform mechanical work, and then construct an appropriate infinitesimal cycle, which
must be reversible in the appropriate limit. Finally, equate the cycle's efficiency to the efficiency
of the corresponding Carnot cycle.

\appendix
\section{Derivation from the Gibbs free energy}\label{Appendix-A}
We now show that the method described above yields correct results in general. Unlike the rest
of the paper, we appeal to concepts beyond the high school program.

Let $X$ be an extensive (property such as volume, or total magnetic moment in the superconducting example) and $Y$ an intensive (such as temperature) parameter characterizing the system under
consideration. The first law of thermodynamics takes the form
\begin{equation}
	dU = \dbar Q - \dbar W = TdS - YdX ~\!, \label{eq-appendix-dU}
\end{equation}
($\dbar Q$ the heat \textit{received} and $\dbar W$ is the work performed \textit{by} the system). The corresponding change in the Gibbs free energy is
\begin{equation}
	dG = -SdT + XdY ~\!.
\end{equation}
(If the system is characterized by some other parameters, assume that they are held constant
during the phase transition that we are considering.) Imagine two infinitesimal processes that
straddle the phase boundary. Since the chemical potentials (Gibbs free energy per particle) of
the two phases (call them $A$ and $B$) are equal at the transition point,
\begin{equation}
	 (S_A - S_B) dT = (X_A - X_B) dY ~\!.
\end{equation}
The latent heat $L= T(S_A - S_B) = T\Delta S$; therefore, on the line of coexistence of both phases
\begin{equation}
\frac{dT}{dY}  = T \frac{\Delta X}{L}.
\label{eq-Appdx-Cl-Cl}
\end{equation}
This is the standard (and the simplest) way to derive the Clausius-Clapeyron relation. Since $YdX$ is the infinitesimal amount of work performed by the system, one can construct a cycle
analogous to the ones considered in the main text. For definiteness assume that in the transition $A\rightarrow B$ the latent heat and $\Delta X$ are both positive, and consider the following steps:
\begin{enumerate}
	\item Transition $A\rightarrow B$ at a temperature $T_0$. The heat absorbed is $L$ and the performed work is $Y\Delta X_{A\rightarrow B}$.
	\item Change $Y \rightarrow Y-\Delta Y$. Note that at this point we already know (because we have assumed that $\Delta X>0$) that the temperature of the phase transition must drop. Otherwise, it would be possible to violate the Kelvin's principle.
	\item Using the colder reservoir, change the temperature of the working substance to $T_0 - \Delta T$ to cause the phase transition $B\rightarrow A$. This is accompanied by the performance by the system of the negative work $-\Delta X_{B\rightarrow A} (Y - \Delta Y)$.
	\item Change the parameter $Y$ and heat up the working substance bringing it back to the initial state.
\end{enumerate}
If $\Delta Y$ and $\Delta T$ are infinitesimal, $\Delta X_{B\rightarrow A}=\Delta X_{B\rightarrow A}$, and one can neglect the heat needed to warm up the working substance in  step 4. The useful work obtained is $\Delta X \Delta Y$, while the heat absorbed is $L$. Comparison with the Carnot engine efficiency yields the desired formula (Eq. \ref{eq-Appdx-Cl-Cl}).
\begin{figure}[H]
    \centering
    \begin{tikzpicture}[scale = 0.9]
	\draw[black, thick, fill=purple!40!pink] (0,0) circle (0.9);
	\node at (0,0) {$A, X$};
	\node at (0, 1.5) {$T_0 - \Delta T, Y$};
	\draw [->] (1.5, 0) -- (6.2, 0);
	\node at (4, 0.5) {$1$};

	\draw[black, thick, fill=green!30!yellow] (8,0) circle (1.2);
	\node at (8,0) {$B, X+\Delta X$};
	\node at (8, 1.5) {$T_0, Y$};
	\draw [->] (8, -1.5) -- (8, -2.5);
	\node at (8.5, -2) {$2$};

	\draw[black, thick, fill=green!30!yellow] (8,-4) circle (1.2);
	\node at (8,-4) {$B, X+\Delta X$};
	\node at (8,-5.5) {$T_0, Y - \Delta Y$};
	\draw [<-] (0, -1.5) -- (0, -2.5);
	\node at (4, -3.5) {$3$};

	\draw[black, thick, fill=purple!40!pink] (0,-4) circle (0.9);
	\node at (0,-4) {$A, X$};
	\node at (0, -5.5) {$T_0 - \Delta T, Y - \Delta Y$};
	\draw [<-] (1.5, -4) -- (6.2, -4);
	\node at (0.5, -2) {$4$};

	\draw [red, thick, -{Stealth}]  (3, 1.5) -- (3, 0.3);
	\node at (3.3, 0.9) {$\color{red}L$};

    \end{tikzpicture}
    \caption{A graphical representation of steps described above. Different colors and sizes of circles representing the working body denote distinct phases, but not necessarily different volumes (the extensive parameter which changes discontinuously during the phase transition is not specified). In step 1, using the heat $L$ one obtains the phase transition $A\rightarrow B$, which results in a change of some extensive parameter $X$, so that work $Y\Delta X$ is performed. In step 2, one changes the intensive parameter $Y$, and as a result the temperature of phase transition drops.  In step 3, we lower the temperature of the body, so that there is another phase transition $B\rightarrow A$, and a negative work $(Y - \Delta Y)\Delta X$ performed. Finally, in step 4 one restores the parameter $Y$ to its initial value.}
    \label{fig-ann}
\end{figure}
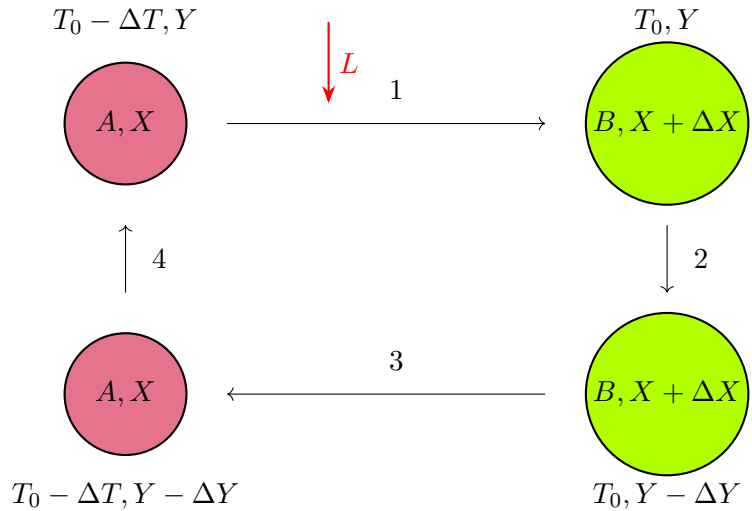

\begin{acknowledgments}

This work is an effect of the remote workshop in physics of the
Polish Children’s Fund. We would like to thank Prof. Piotr Chankowski for his suggestions
and invaluable help.

\end{acknowledgments}

\end{document}